\documentclass{article}
\usepackage[utf8]{inputenc}
\usepackage{amsmath}
\usepackage{amssymb}
\usepackage{xcolor}

\usepackage{authblk}

\begin{document}
\title{Thermodynamic conditions ensure the stability of third-order extended heat conduction}
\author[1,2,4]{P. Ván}
\author[1,3,4]{R. Somogyfoki}

\affil[1]{Department of Theoretical Physics, Institute of Particle and Nuclear Physics, HUN-REN Wigner Research Centre for Physics, Konkoly Thege Miklós út. 29-33, Budapest H-1525, Hungary}
\affil[2]{Department of Energy Engineering, Faculty of Mechanical Engineering, Budapest University of Technology and Economics,
	Műgyetem rkp. 3, Budapest 1111, Hungary}
\affil[3]{ELTE Doctoral School of Physics, Eötvös Loránd University, Pázmány Péter stny. 1/A, Budapest 1117, Hungary}
\affil[4]{Montavid Thermodynamic Research Group, Budapest, Hungary}

\maketitle

\begin{abstract}
In a recent work, Somogyfoki et al. (J. Non-Equilib. Thermodyn. 50, 59-76, 2025) \cite{1} analysed the linear stability of homogeneous equilibrium in third-order non-Fourier heat conduction within the framework of non-equilibrium thermodynamics with internal variables. They identified a stability condition, their equation (49), which could not be derived from the standard thermodynamic inequalities for the $2\times2$ conductivity blocks, and concluded that the Second Law does not guarantee stability in the most general case. \textcolor{black}{Here we show that this conclusion was due to an overly conservative proof strategy:} the standard thermodynamic conditions (concave entropy and non-negative entropy production, as expressed by the $2\times2$ block positive-definiteness inequalities (19)-(20) of the original paper) do suffice for linear stability. \textcolor{black}{The key observation is that all coefficients of the dispersion polynomial remain positive for all physical wave numbers because their structure prevents positive real roots.} This result confirms that thermodynamics, understood as a stability theory \cite{2}, ensures fundamental dynamic stability in all thermodynamically consistent third-order extended heat conduction theories. A comparison with the rate-equation approach of Giorgi, Morro and Zullo (Meccanica 59, 1757-1776, 2024) \cite{8} is also presented.
\end{abstract}

\section{Introduction}
The relationship between thermodynamics and stability has a long and fruitful history. In a recent paper \cite{2}, it was argued that thermodynamics is, at its core, a stability theory: the postulates of classical thermodynamics (the existence of entropy as a potential, its concavity, and the non-negativity of entropy production) are precisely the conditions for the asymptotic Lyapunov stability of thermodynamic equilibrium. A neceessary condition is the linear stability of the equilibrium. This perspective was demonstrated for ordinary (homogeneous) thermodynamics in \cite{2,3}, and for the Fourier-Navier-Stokes system in \cite{4}.

A natural testing ground for this thesis is the stability analysis of higher-order continuum theories. Somogyfoki et al. \cite{1} \textcolor{black}{performed} such an analysis for third-order extended theories of heat conduction in one spatial dimension, using the framework of Non-equilibrium Thermodynamics with Internal Variables (NET-IV) \cite{5,6}. Their general conclusion was optimistic: thermodynamic conditions ensure stability in all physically relevant special cases (Fourier, Maxwell-Cattaneo-Vernotte, Guyer-Krumhansl, Extended Thermodynamics, and several others). However, they identified a condition, equation (49) in \cite{1}:
\begin{equation}\label{tcond}
\lambda_{11}\lambda_{22}+\kappa_{11}\kappa_{22}-(\hat{\lambda}-\hat{\kappa})^{2}+(\check{\lambda}-\check{\kappa})^{2}\ge0 \tag{1}
\end{equation}
which, while satisfied in all known special cases, could not be derived from the standard thermodynamic inequalities at the most general level. \textcolor{black}{In the present paper, we show that condition (1) is in fact too strict: the standard thermodynamic conditions alone are sufficient for all Routh-Hurwitz stability requirements.}

\section{Setup}
We adopt the notation of \cite{1}. The one-dimensional linearised system for perturbations $(\delta T, \delta q, \delta Q)$ around equilibrium $(T_{0},0,0)$ is governed by the $4\times4$ conductivity matrix:
\begin{equation}
L=\begin{pmatrix}\lambda_{1}&0&0&\lambda_{12}\\ 0&\kappa_{1}&\kappa_{12}&0\\ 0&\kappa_{21}&\kappa_{22}&0\\ \lambda_{21}&0&0&\lambda_{2}\end{pmatrix} \tag{2}
\end{equation}
The non-negative entropy production conditions reduce to the two $2\times2$ block requirements:
\begin{equation}
\lambda_{1}\ge0, \quad \lambda_{2}\ge0, \quad \lambda_{1}\lambda_{2}\ge\hat{\lambda}^{2} \tag{3}
\end{equation}
\begin{equation}
\kappa_{1}\ge0, \quad \kappa_{2}\ge0, \quad \kappa_{1}\kappa_{2}\ge\hat{\kappa}^{2} \tag{4}
\end{equation}
where $\hat{\lambda}=(\lambda_{12}+\lambda_{21})/2$ and $\hat{\kappa}=(\kappa_{12}+\kappa_{21})/2$. The dispersion polynomial for exponential plane waves $\propto e^{\Gamma t+ikx}$ is cubic in $\Gamma$:
\begin{equation}
a_{0}\Gamma^{3}+a_{1}\Gamma^{2}+a_{2}\Gamma+a_{3}=0 \tag{5}
\end{equation}
with coefficients (writing $X=k^{2}$ for brevity):
\begin{align*}
a_{0} &= mM \tag{6} \\
a_{1} &= (M\kappa_{1}+m\lambda_{2})X+m\kappa_{2}+M\lambda_{1} \tag{7} \\
a_{2} &= \kappa_{1}\lambda_{2}X^{2}+(M\alpha+Z)X+\kappa_{2}\lambda_{1} \tag{8} \\
a_{3} &= \alpha X(\kappa_{2}+\lambda_{2}X) \tag{9}
\end{align*}
where $\alpha=M/(\rho c_{v}T_{0}^{2}) > 0$. \textcolor{black}{Note that $\alpha$ is defined by the inductivity $M$ and specific heat $c_v$, which are independent of the conductivity block parameters.} The coefficient $Z$ is the expression in (\ref{tcond}).
$$
Z=\kappa_{1}\kappa_{2}+\lambda_{1}\lambda_{2}-(\kappa_{12}-\lambda_{12})(\kappa_{21}-\lambda_{21})=
\kappa_{1}\kappa_{2}+\lambda_{1}\lambda_{2}-(\hat\kappa-\hat\lambda)^2 +(\check\kappa-\check\lambda)^2,
$$
where $\check{\lambda}=(\lambda_{12}-\lambda_{21})/2$ and $\check{\kappa}=(\kappa_{12}-\kappa_{21})/2$, the antisymmetric coefficients of the block matrices.
The Routh-Hurwitz conditions for all roots to have negative real parts are: $a_{j}>0$ for $j=0,1,2,3$ and $a_{1}a_{2}>a_{0}a_{3}$.

\section{Main result}
\textbf{Theorem 1.} If $\alpha>0$, the thermodynamic inductivities satisfy $m,M>0$, \textcolor{black}{and the conductivity blocks satisfy the strict thermodynamic inequalities $\lambda_1\lambda_2 > \hat{\lambda}^2$ and $\kappa_1\kappa_2 > \hat{\kappa}^2$}, then for every $k\ne0$, all roots of (5) have strictly negative real parts.

\textbf{Proof.} \\
\textbf{Step 1: $a_{0}, a_{1}, a_{3} > 0$}. This is straightforward : $a_{0}=mM>0$; $a_{1} > 0$ since all terms are non-negative and the constant part $m\kappa_{2}+M\lambda_{1}$ is positive; $a_{3}=\alpha X(\kappa_{2}+\lambda_{2}X)>0$ for $X>0$. \\
\textbf{Step 2: $a_{2} > 0$}. \textcolor{black}{Since the $M\alpha X$ term in $a_{2}$ is strictly positive, it suffices to prove the positivity of $f(X)=\kappa_{1}\lambda_{2}X^{2}+ZX+\kappa_{2}\lambda_{1}$ for all $X>0$.} A lower bound of $f(X)$ is $g(X) = \kappa_1\lambda_2 X^2 + (A+B-w^2)X + \kappa_2\lambda_1$, where $A=\kappa_1\kappa_2$, $B=\lambda_1\lambda_2$, and $w=\hat{\kappa}-\hat{\lambda}$. \begin{itemize}
    \item \textcolor{black}{If $A+B\ge w^2 $, then by Descartes' Rule of Signs, there are no positive roots as all coefficients are non-negative.}
    \item \textcolor{black}{If $A+B < w^2  < 0$, a positive root would require the discriminant 
    $$
    \Delta = (A+B-w^2)^2-4AB =-((\sqrt{A}-\sqrt{B})^2+w^2)((\sqrt{A}+\sqrt{B})^2-w^2)
    $$
    to be non-negative. However, the block conditions force $|w| < \sqrt{A}+\sqrt{B}$, which strictly implies $\Delta < 0$.}
\end{itemize}
\textcolor{black}{In both cases, $a_2 > 0$.} \\
\textbf{Step 3: $a_{1}a_{2}-a_{0}a_{3}>0$}. The expression simplifies to $a_{1}f(X)+\alpha X[M^{2}(\lambda_{1}+\kappa_{1}X)]$. Since $f(X)>0$ and all other parameters are positive, the condition is satisfied.

\section{Discussion}
\subsection{Thermodynamics as a stability theory}
Theorem 1 provides a clear confirmation of the thesis in \cite{2}: the standard thermodynamic conditions are sufficient for the linear stability of homogeneous equilibrium. \textcolor{black}{Asymptotic stability is guaranteed specifically by the strict positive-definiteness of the conductivity blocks.}

\subsection{\textcolor{black}{Mechanism of Stability}}
\textcolor{black}{The stability of $a_2$ is ensured even when $Z < 0$ because the thermodynamic bounds on cross-coupling ($|w|$) are tighter than the bounds required for the discriminant to become positive. This makes condition (49) of \cite{1} sufficient but not necessary.}

\subsection{Comparison with the rate-equation approach of Giorgi--Morro--Zullo}
Giorgi, Morro and Zullo \cite{8,GioZul26a} developed a systematic approach to heat conduction through rate-type constitutive equations derived via the Coleman--Noll procedure. Their method prescribes the entropy production as a non-negative constitutive function ab initio. In their framework, several heat conduction models are recovered as special cases: the Fourier, Maxwell--Cattaneo--Vernotte, Jeffreys, Green--Naghdi, and Quintanilla models. 

The key structural difference from the NET-IV approach \cite{1,5} lies in the entropy flux: NET-IV employs generalised entropy current densities with current multipliers \cite{9}, which allow coupling between thermodynamic forces and fluxes of different tensorial orders. This generates the full cross-coefficient structure of the conductivity matrix (2). \textcolor{black}{While the Coleman-Noll approach requires case-by-case verification, Theorem 1 in the NET-IV framework provides a blanket guarantee: any thermodynamically consistent third-order model is automatically stable.}

\subsection{Outlook}
Several extensions are worth exploring. The three-dimensional generalisation, where isotropic representation theory introduces additional complications \cite{7}, is expected to follow similar logic but requires separate analysis. Pure internal variable approach, with a vectorial internal variable in the state space instead of the heat flux, is expected to resolve the peculiar degeneracies of the Extended Thermodynamic framework. Higher-order theories (fourth-order and beyond), relevant for relativistic fluid dynamics \cite{10}, are a natural next step. Finally, the connection between linear stability and full nonlinear Lyapunov stability remains an important open problem.

\section{Conclusions}
The standard thermodynamic conditions (concave entropy and non-negative entropy production) expressed through the \textcolor{black}{strict} block positive-definiteness inequalities (3)-(4) are sufficient for the linear stability of third-order extended heat conduction. The intermediate stability condition (49) of \cite{1} is therefore too strict. The unconditional, sharpened proof confirms the thesis of \cite{2}: thermodynamics is \textcolor{black}{indeed} a stability theory, and fundamental dynamic stability is a universal consequence of the Second Law.

\section{Acknowledgement}

The author acknowledge networking support by the grant NKFIH NKKP-Advanced 150038 and COSTAction FuSe CA24101. This article/publication is based upon work from COST Action FuSe, CA24101, supported by COST (European Cooperation in Science and Technology).

The authors used several AI assistants for editing and brainstorming.

\end{document}